\documentstyle[12pt,epsfig]{article}

\newcommand{\lsim}{\raisebox{-0.13cm}{~\shortstack{$<$ \\[-0.07cm] $\sim$}}~}

\newcommand{\ra}{\rightarrow}

\newcommand{\s}{\\ \vspace*{-3mm} }
\newcommand{\nn}{\noindent}
\newcommand{\non}{\nonumber}
\newcommand{\beq}{\begin{eqnarray}}
\newcommand{\eeq}{\end{eqnarray}}

\newcommand{\tb}{\tan\beta}

\begin{document}

\def\thefootnote{\fnsymbol{footnote}}

\begin{flushright}
KA--TP--14--96\\
May 1996 \\
\end{flushright}

\vspace{1cm}

\begin{center}

{\large\sc {\bf QCD Corrections to the Top}}

\vspace*{3mm}

{\large\sc {\bf Decay Mode $t \ra \tilde{t} \chi^0$}}

\vspace{1cm}

{\sc A.~Djouadi\footnote{Supported by Deutsche Forschungsgemeinschaft
DFG (Bonn).}, W. Hollik,} and {\sc C. J\"unger}

\vspace{1cm}

Institut f\"ur Theoretische Physik, Universit\"at Karlsruhe,\\
\vspace{0.2cm}
D--76128 Karlsruhe, Germany. 

\end{center}

\vspace{2cm}

\begin{abstract}

\nn In supersymmetric theories, the top quark can decay into its scalar
partner plus a neutralino, with an appreciable rate. We calculate the
${\cal O}(\alpha_s)$ QCD corrections to this decay mode in the minimal 
supersymmetric extension of the Standard Model. These corrections can 
be either positive or negative and increase logarithmically with the 
gluino mass. For gluino masses below 1 TeV, they are at most of the 
order of ten percent and therefore, well under control. 

\end{abstract}

\newpage

\def\thefootnote{\arabic{footnote}}
\setcounter{footnote}{0}

\subsection*{1. Introduction}

It is well--known that in supersymmetric theories (SUSY), the heavy top 
quark can decay at tree--level into its lightest scalar partner and 
a light neutralino state \cite{R1,R2}. This possibility is intimately 
related to the large value of the top mass $m_t$: the mixing between the 
left-- and right--handed scalar partners of the top quark, $\tilde{t}_L$ 
and $\tilde{t}_R$, is proportional to $m_t$; after diagonalization of the 
mass matrix, the lightest stop mass eigenstate $\tilde{t}$ can be much 
lighter than the top quark and all the scalar partners of the light quarks 
\cite{R1}. \s

The experimental bounds \cite{R3} on the lightest stop and neutralino
masses leave a large room for this decay to occur. Moreover, light stop
squarks and chargino states [and hence the light neutralino 
$\chi_1^0$ which is the
lightest supersymmetric particle] might explain the large deviation of
the experimentally measured value of the ratio $R_b =\Gamma(Z \ra
b\bar{b})/ \Gamma (Z \ra {\rm hadrons})$ from the Standard Model
prediction \cite{R4}. In fact, in the area of the MSSM parameter space
which provides the best fit \cite{R4} to the electroweak precision data,
all four neutralinos are light enough [while the charginos are heavier
than the present experimental bounds \cite{R3} from LEP1.5] for the top
quark to decay into the lightest stop squark plus one of the four
neutralino states. \s 

In the minimal supersymmetric extension of the Standard Model (MSSM),
the decay $t \ra \tilde{t} \chi_1^0$ could have a sizeable branching
fraction\footnote{The present agreement between the top quark production
cross section at the Tevatron \cite{R6} and the Standard Model
prediction still allows appreciable $t \ra \tilde{t} \chi_1^0$ decay
rates. A detailed analysis of the area of the parameter space where
these decays are allowed or disfavored by Tevatron data is beyond the
scope of the present paper.} when it is allowed kinematically \cite{R2}.
In a large area of the SUSY parameter space, this decay can have a
signature similar to that of the standard decay $t \ra bW^+ \ra b l^+
\nu$ or $b$+2jets \cite{R5}. Indeed, the stop could mainly decay into
the lightest chargino and a bottom quark, and the chargino will
subsequently decay into a neutralino and a virtual $W$ boson which then
goes into $l^+ \nu$ or 2--jets. This gives the same topology as the $ t
\ra bW^+$ channel, except for the large amount of missing energy in the
SUSY decay mode due to the escaping neutralinos, which leads to a softer
transverse momentum spectrum of the charged leptons. Such decays could
have an impact on the top quark studies at the Tevatron \cite{R6},
altering for instance the reported production cross section
values\footnote{For such light stop squarks, the pair production $pp \ra
\tilde{t} \tilde{t}^*$ is possible at the Tevatron and will also
contribute to the event samples; for a recent discussion of this
possibility see for instance Ref.\cite{R7}.} since these events would not
all pass the experimental cuts designed for the standard $t\ra b W^+$
decay. \s 

A detailed study of the $t \ra \tilde{t} \chi^0_1$ decay mode is
therefore mandatory. To have a more reliable prediction for the decay
branching ratio, the QCD corrections have to be taken into account. The
one--loop QCD corrections to the two other decay modes of the top quark
in supersymmetric theories, $ t \ra bW^+$ and $ t \ra H^+b$, can be
found in Refs.\cite{R8} and \cite{R9} respectively. In this article, we present the
${\cal O}(\alpha_s)$ QCD corrections to the top decay into a stop squark
and a neutralino. These results can be easily generalized to the 
possible decay into a light sbottom squark and a chargino state. 

\subsection*{2. Born Approximation}

In the MSSM, the decay width $\Gamma(t \ra \tilde{t}_i \chi_j^0$) [where
$i=1,2$ are for the stop and $j=1$--$4$ for the neutralino states] will
depend on five parameters: the left-- and right--handed scalar stop
masses [which in general are taken to be equal], the Higgs--higgsino
mass parameter $\mu$, the soft--SUSY breaking trilinear coupling $A_t$,
the ratio of the vacuum expectation values of the two Higgs doublet MSSM
fields $\tb$ [which fix the stop masses and the mixing angle] as well as
the wino mass parameter $M_2$ [which together with $\mu$ and $\tb$ fixes
the neutralino masses]; the gluino mass is related to the parameter $M_2,
m_{\tilde{g}} \sim 3.5 M_2$, when the gaugino masses and the three
coupling constants of SU(3)$\times$SU(2)$\times$U(1) are unified at the
Grand Unification scale. \s 

In the Born approximation, the partial widths for the decays $t \ra
\tilde{t}_i \chi_j^0$ [we will drop for simplicity the indices for the 
neutralino masses, $m_{\chi} \equiv m_{\chi_j^0}$] are given by 
\begin{equation}
\Gamma(t \rightarrow \tilde{t}_i \chi_j^0 )=  \frac{\alpha}
{8 \, m_t^3} \; 
      \bigg\{ a_L^i \,a_R^i \, ( 4 \, m_t \, m_{\chi}\epsilon_\chi )
    + ({a_L^i}^2 + {a_R^i}^2) \,  ( m_t^2 - m_{\tilde{t}_i}^2 + m_{\chi}^2 )
      \bigg\}\,\lambda^{1/2}(m_t^2,m_{\chi}^2,m_{\tilde{t}_i}^2)
\end{equation}
where $\lambda (x,y,z)=x^2+y^2+z^2-2\,(xy+xz+yz)$ is the usual 
two--body phase space function and $\epsilon_\chi$ is the sign of the
eigenvalue of the neutralino $\chi$; the couplings $a^i_{L,R}$ which
form the Born $\chi t \tilde{t}_i$  vertex 
\beq
\Gamma_0^i = ie \left[a_{L}^i P_L + a_R^i P_R \right]
\eeq
with $P_{L,R}$ the chirality projectors $P_{L,R}=(1 \mp \gamma_5)/2$,
are given by
\begin{displaymath}
\left\{ \begin{array}{c} a_L^1 \\ a_L^2 \end{array} \right\} =
   b \, m_t \, 
\left\{ \begin{array}{c} s_{\theta} \\ c_{\theta} \end{array} \right\}
 + \, c_L \,
\left\{ \begin{array}{c} c_{\theta} \\-s_{\theta} \end{array} \right\}
\; , \mbox{\hspace{1.cm}}
\left\{ \begin{array}{c} a_R^1 \\ a_R^2 \end{array} \right\} =
   b \, m_t \, 
\left\{ \begin{array}{c} c_{\theta} \\-s_{\theta} \end{array} \right\}
 + \, c_R \,
\left\{ \begin{array}{c} s_{\theta} \\ c_{\theta} \end{array} \right\}
\end{displaymath}
\begin{eqnarray}
b & = & \frac{1}{\sqrt{2}\, M_W \sin\beta s_W} \; N_{j4} \nonumber \\ 
c_L & = & \sqrt{2}\left[ \frac{2}{3} \; N_{j1}'
              + \left(\frac{1}{2} - \frac{2}{3}\, s_W^2 \right)
                 \frac{1}{c_W s_W}\;N_{j2}' \right]  \nonumber \\ 
c_R & = &-\sqrt{2}\left[ \frac{2}{3} \; N_{j1}'
              - \frac{2}{3} \frac{s_W}{c_W} \; N_{j2}' \right]
\end{eqnarray}
Here $\theta$ is the stop mixing angle [which can be expressed in terms of
the Higgs--higgsino SUSY mass parameter $\mu$, $\tb$ and the soft--SUSY
breaking trilinear coupling $A_t$] with $s_{\theta}=\sin\theta$,
$c_{\theta}=\cos\theta$ etc.; $s_W^2=1-c_W^2\equiv \sin^2\theta_W$.
$N$ is the diagonalizing matrix for the neutralino states \cite{R10} and 
\beq
N'_{j1}= c_W N_{j1} +s_W N_{j2} \ \ \ , \ \ \
N'_{j2}= -s_W N_{j1} +c_W N_{j2} \ \ 
\eeq
The masses of the two stop squarks and their
mixing angle are determined by diagonalizing the following mass matrix:
\begin{equation}
{\cal M}^2_{\tilde{t}} = 
\left( 
  \begin{array}{cc} M_{\tilde{t}_L}^2 + m_t^2 + \cos 2 \beta (\frac{1}{2}
                       - \frac{2}{3}s_W^2) \, M_Z^2  & m_t \, M_{LR} \\
                    m_t \, M_{LR} & M_{\tilde{t}_R}^2 + m_t^2
                                   + \frac{2}{3}\cos 2 \beta \; s_W^2 \, M_Z^2
  \end{array}
\right)
\end{equation}
where $M_{LR}$ in the off--diagonal term  reads $M_{LR} = A_t - \mu \, \cot 
\beta$. \s

Assuming that the MSSM charged Higgs boson, $H^+$, is heavier than the
top quark so that the decay channel $t \ra H^+ b$ is kinematically
closed, the branching ratios for the $t \ra \tilde{t}_2 \chi_j^0$ decay
modes [$\tilde{t}_2$ is the lightest stop squark in the notation
introduced above] are shown in Fig.~1 for $\tb=1.6$, a value favored by
theories with $b$--$\tau$ Yukawa coupling unification. The top quark 
mass is fixed to $m_t = 180$ GeV. In Fig.1a, we fix
the wino mass to $M_2= \mu =150, 200$ and 250 GeV and plot
the branching ratio for the decay into the stop and the lightest neutralinos as
a function of the $\tilde{t}_2$ mass [obtained by varying the parameter
$A_t$]. In Fig.1b, $M_2$ is kept fixed at $M_2=400$ GeV, and two values have been
chosen for $\mu, -100$ and 150 GeV. Whenever the decay in more than one 
of the neutralinos is possible, the branching ratio has been summed.
As can be seen, it can reach the level of several 
ten percent, and can even dominate over
the standard $t \ra bW^+$ decay mode. \s 

In Fig.1c, we choose the parameter $\mu$ to be negative and
equal to the wino mass $M_2 =-\mu = 50$ GeV; this area of
the MSSM parameter space, together with a rather light right--handed
stop squark, $m_{\tilde{t}} \lsim 100$ GeV, provides the best MSSM fit
to the high--precision electroweak data, and enhances the ratio $R_b
=\Gamma( Z \ra b\bar{b})/ \Gamma (Z \ra {\rm hadrons})$ to the level
where it is almost consistent with its experimental value \cite{R4}. The
branching ratios of the four decays $t \ra \tilde{t}+\chi_j^0$ with
$j=$1--4 [the masses of the neutralinos can be read off from the
thresholds, and the lightest chargino in this scenario has a mass
$m_{\chi_1^+} \sim 70$ GeV] are shown together with their sum. 
They are also large, and can dominate over the $t\ra bW^+$ branching ratio.

\subsection*{3. QCD corrections}

The QCD corrections to the decay width, eq.(1), consist of virtual
corrections Figs.2a--d, and real corrections with an additional gluon
emitted off the initial or final state, Fig.2e. The ${\cal O}(\alpha_s)$ virtual
contributions can be split into gluon and gluino exchange in the
$t$--$\tilde{t}$--$\chi$ vertex, and the top and stop wave function
renormalizations. The renormalization of the $t$--$\tilde{t}$--$\chi$
coupling is achieved by renormalizing the top quark mass and the stop
mixing angle. We will use the dimensional reduction 
scheme \cite{R11} which preserves 
supersymmetry\footnote{Because the $\chi \tilde{t}t$ coupling is renormalized by QCD,
the dimensional regularization and dimensional reduction schemes 
give slightly different results.}  to
regularize the ultraviolet divergencies and we introduce a fictitious gluon
mass $\lambda$ to regularize the infrared divergencies. The sum of the
virtual corrections obtained by taking the interference between the
Born and QCD corrected amplitudes, and the real correction, obtained by
squaring the sum of the two diagrams in Fig.2e, is finite as it should
be. 

\subsubsection*{3.1 Virtual Corrections}

The QCD corrections to the $t$--$\tilde{t}_i$--$\chi$ vertex, eq.(2), 
can be written as
\beq 
\delta \Gamma^i = ie \ \frac{\alpha_s}{3\pi} \,
\sum_{j=g, \tilde{g}, mix, ct} \left[ G_{j,L}^i P_L + G_{j,R}^i P_R \right]
\eeq 
where $G^i_{g}, \,G^i_{\tilde{g}}, \,G^i_{mix}$ and $G^i_{ct}$ denote the 
gluon and gluino exchanges in the vertex, and the mixing and counterterm 
contributions, respectively. \s

The gluon exchange contributions [Fig.2a] can be written as
\begin{eqnarray}
G^i_{g,L} = a_L^i \, F_1^i + a_R^i \,F_2^i \ \ \ , \ \ \
G^i_{g,R} = a_R^i \, F_1^i + a_L^i \,F_2^i 
\end{eqnarray}
with the form factors $F^i_{1,2}$ given by
\beq
F_1^i & = &  B_0(m_{\tilde{t}_i}^2,\lambda,m_{\tilde{t}_i})
 + 2 \, m_t^2 \, C_0 - 2 \, m_{\tilde{t}_i}^2 \, (C_{11}-C_{12}) 
 + 2 \, m_{\chi}^2 \, C_{11} \nonumber \\
F_2^i & = & -2 \, m_t \, m_{\chi} \, (C_0+C_{11})
\end{eqnarray}
where the two and three--point Passarino--Veltman functions, $B_0$ and 
$C_{..} \equiv C_{..}(m_t^2, m_{\tilde{t}_i}^2, m_{\chi}^2,$ $m_t^2, \lambda^2, 
m_{\tilde{t}_i}^2)$ can be found in Ref.\cite{R12}. \s

The gluino exchange contributions [Fig2b], are given by
\beq
G_{\tilde{g},L}^1 & = & 
a_L^1 ( s_{2 \theta} F_4^{11}-c_{2 \theta} F_5^{11}-F_5^{11} )
+ a_R^1( s_{2 \theta} F_1^{11}-c_{2 \theta} F_2^{11}+F_3^{11} )
\nonumber \\ &+ & 
a_L^2( s_{2 \theta} F_5^{12}+c_{2 \theta} F_4^{12}-F_7^{12} )
+ a_R^2( s_{2 \theta} F_2^{12}+c_{2 \theta} F_1^{12}-F_6^{12} )  
\nonumber \\
G_{\tilde{g},R}^1 &=& 
a_L^1 ( s_{2 \theta} F_1^{11}+c_{2 \theta} F_2^{11}+F_3^{11} )
+ a_R^1( s_{2 \theta} F_4^{11}+c_{2 \theta} F_5^{11}-F_5^{11} )
\nonumber \\   & + & 
a_L^2(-s_{2 \theta} F_2^{12}+c_{2 \theta} F_1^{12}+F_6^{12} )
+ a_R^2(-s_{2 \theta} F_5^{12}+c_{2 \theta} F_4^{12}+F_7^{12} )
\nonumber \\
G_{\tilde{g},L}^2 & = & 
a_L^1 ( s_{2 \theta} F_5^{21}+c_{2 \theta} F_4^{21}+F_7^{21} )
+ a_R^1 ( s_{2 \theta} F_2^{21}+c_{2 \theta} F_1^{21}+F_6^{21} )
\nonumber \\
& + & 
a_L^2 (-s_{2 \theta} F_4^{22}+c_{2 \theta} F_5^{22}-F_5^{22} )
+ a_R^2 (-s_{2 \theta} F_1^{22}+c_{2 \theta} F_2^{22}+F_3^{22} )
\nonumber \\ 
G_{\tilde{g},R}^2 & = &  
a_L^1 (-s_{2 \theta} F_2^{21}+c_{2 \theta} F_1^{21}-F_6^{21} )
+ a_R^1 (-s_{2 \theta} F_5^{21}+c_{2 \theta} F_4^{21}-F_7^{21} )
\nonumber \\
& + & 
a_L^2 (-s_{2 \theta} F_1^{22}-c_{2 \theta} F_2^{22}+F_3^{22} )
+ a_R^2 (-s_{2 \theta} F_4^{22}-c_{2 \theta} F_5^{22}-F_5^{22} )  
\eeq
where in terms of the functions
$C_{..}^{ik}=C_{..}^{ik}(m_t^2,m_{\tilde{t}_i}^2,m_{\chi},
m_{\tilde{t}_k}^2,m_{\tilde{g}}^2,m_t^2)$ [$k$ denotes the 
virtual (summed over) and $i$ the outgoing squarks], the form 
factors $F^{ik}$ are given by
\beq
F_1^{ik} & = & (m_{\tilde{t}_k}^2 + m_t^2) \, C_0^{ik} 
                + 2 \, m_t^2 \, C_{11}^{ik}
       + (m_{\chi}^2 - 2 \, m_t^2) \, C_{12}^{ik}  
       + B_0(m_{\tilde{t}_i}^2,m_{\tilde{g}},m_t)
  \nonumber \\
F_2^{ik} & = & m_{\tilde{g}} \, m_t \, (C_0^{ik}
                          - C_{11}^{ik}
                          + C_{12}^{ik})  \nonumber \\
F_3^{ik} & = & m_{\tilde{g}} \, m_t \, ( - C_0^{ik} 
                          - C_{11}^{ik}
                          + C_{12}^{ik} )  \nonumber \\
F_4^{ik} & = & m_t \, m_{\chi} \, ( C_0^{ik} 
                          + C_{11}^{ik}  
                          + C_{12}^{ik} ) \nonumber \\
F_5^{ik} & = & m_{\tilde{g}} \, m_{\chi} \, ( C_0^{ik} 
                          + C_{12}^{ik})  \nonumber \\
F_6^{ik} & = & (m_{\tilde{t}_k}^2 - m_t^2) \, C_0^{ik}
          + m_{\chi}^2 C_{12}^{ik} 
          + B_0(m_{\tilde{t}_i}^2,m_{\tilde{g}},m_t)
  \nonumber \\
F_7^{ik} & = & m_t \, m_{\chi} ( - C_0^{ik}
                          - C_{11}^{ik} 
                          + C_{12}^{ik} )
\eeq 

The mixing contribution is due to the diagrams of Fig.2c where the
top--gluino exchange and the quartic scalar interaction contributions
switch the Born $\chi t \tilde{t}_1$ vertex to the ${\cal O}(\alpha_s)$
$\chi t \tilde{t}_2$ one and {\it vice versa}; it is given by 
\begin{eqnarray}
G_{mix,L}^i & = & \frac{(-1)^i\,(\delta_{1i}\,a_L^2 + \delta_{2i}\,a_L^1)}
                       {m_{\tilde{t}_1}^2-m_{\tilde{t}_2}^2} \,
\left[ 4 m_t \,m_{\tilde{g}}\, c_{2 \theta}\,B_0(m_{\tilde{t}_i}^2,
      m_t,m_{\tilde{g}})   + c_{2 \theta} s_{2\theta} (
A_0(m_{\tilde{t}_2}^2)-  A_0(m_{\tilde{t}_1}^2) ) \right] \nonumber \\
G_{mix,R}^i & = & \frac{(-1)^i\,(\delta_{1i}\,a_R^2 + \delta_{2i}\,a_R^1)}
                       {m_{\tilde{t}_1}^2-m_{\tilde{t}_2}^2} \,
\left[ 4     m_t \,m_{\tilde{g}}\, 
c_{2 \theta}\,B_0(m_{\tilde{t}_i}^2,m_t,m_{\tilde{g}})
+ c_{2 \theta} s_{2\theta} (A_0(m_{\tilde{t}_2}^2)- 
A_0(m_{\tilde{t}_1}^2) ) \right] \nonumber \\
&& 
\end{eqnarray}

\subsubsection*{3.2 Counterterms}

The counterterm contributions in eq.(6) are due to the top and stop wave
function renormalizations [Fig.2d] as well as the renormalization of the
top quark mass $m_t$ and the mixing angle $\theta$, which appear in the
Born coupling, 
\begin{eqnarray}
G_{ct,L}^{1,2} & = & \frac{1}{2} a_L^{1,2} \left( \delta Z^t_L 
+ \delta Z_{\tilde{t}_{1,2}} \right)
              + b \, \{s_{\theta},c_{\theta} \} \,\delta m_t
              + b \,m_t \, \{c_{\theta},-s_{\theta}\} \,\delta\theta
              - c_L \, \{s_{\theta},c_{\theta}\} \,\delta\theta
   \nonumber \\
G_{ct,R}^{1,2} & = & \frac{1}{2} a_R^{1,2} \left( \delta Z^t_R
+ \delta Z_{\tilde{t}_{1,2}} \right)
              + b \,\{c_{\theta},-s_{\theta}\} \,\delta m_t
              - b \,m_t \, \{s_{\theta},c_{\theta}\} \,\delta\theta
              + c_R \, \{c_{\theta},-s_{\theta} \} \,\delta\theta \ \ 
\end{eqnarray}

In the on--shell scheme, the quark and squark masses are defined as 
the poles of the propagators and the wave--function renormalization
constants  follow from the residues at the poles; the corresponding 
counterterms are given by
\beq 
\frac{\delta m_t}{m_t} & = & \frac{1}{2}
\bigg[ \Sigma^t_R(m_t^2)+\Sigma^t_L(m_t^2)\bigg] 
+ \Sigma^t_S(m_t^2)   \nonumber \\
\delta Z^t_L & = & - \Sigma^t_L(m_t^2)
                   - m_t^2 \bigg[ {\Sigma^t_L}^{\prime}(m_t^2)
                   +{\Sigma^t_R}^{\prime}(m_t^2)+2\,{\Sigma^t_S}^{\prime}(m_t^2)
                           \bigg]   \nonumber \\
\delta Z^t_R & = & - \Sigma^t_R(m_t^2) 
                   - m_t^2 \bigg[ {\Sigma^t_L}^{\prime}(m_t^2)
                   +{\Sigma^t_R}^{\prime}(m_t^2)+2\,{\Sigma^t_S}^{\prime}(m_t^2)
                           \bigg]   \nonumber \\
\delta Z_{\tilde{t}_i} & = & - \left(\Sigma_{\tilde{t}}^{ii}
\right)'(m_{\tilde{t}_i}^2)
\eeq
where the self--energies $\Sigma$ and their derivatives $\Sigma'$, up to 
a factor $\alpha_s /3\pi$ which has been factorized out, are given 
in the dimensional reduction scheme by
\beq
\Sigma^t_L(k^2) & = &  - \bigg[ 2 \,B_1(k^2,m_t,\lambda)  
            + (1+c_{2 \theta}) B_1(k^2,m_{\tilde{g}},m_{\tilde{t}_1})
            + (1-c_{2 \theta}) B_1(k^2,m_{\tilde{g}},m_{\tilde{t}_2}) \bigg]
   \nonumber \\
\Sigma^t_R(k^2) & = &  - \bigg[ 2 \,B_1(k^2,m_t,\lambda) 
            + (1-c_{2 \theta}) B_1(k^2,m_{\tilde{g}},m_{\tilde{t}_1})
            + (1+c_{2 \theta}) B_1(k^2,m_{\tilde{g}},m_{\tilde{t}_2}) \bigg]
   \nonumber \\
\Sigma^t_S(k^2) & = &  -    \bigg[ 4 \,B_0(k^2,m_t,\lambda) 
      + \frac{m_{\tilde{g}}}{m_t}\,s_{2 \theta}\,
         ( B_0(k^2,m_{\tilde{g}},m_{\tilde{t}_1})
         - B_0(k^2,m_{\tilde{g}},m_{\tilde{t}_2}) ) \bigg]
   \nonumber \\
(\Sigma_{\tilde{t}}^{ii})'(k^2) & = & - 2 \bigg[
-2\,B_1(k^2,m_{\tilde{t}_i},\lambda)
-2\,k^2\,B_1'(k^2,m_{\tilde{t}_i},\lambda)  
+ (m_t^2+m_{\tilde{g}}^2-k^2)\,B_0'(k^2,m_t,m_{\tilde{g}}) \nonumber \\
&& \ \ \ - \,B_0(k^2,m_t,m_{\tilde{g}})
         + (-1)^i\,2\,s_{2 \theta}\,m_t\,m_{\tilde{g}} 
B_0'(k^2,m_t,m_{\tilde{g}}) \bigg]
\end{eqnarray}

Finally, the mixing angle counterterm is chosen\footnote{Had we chosen the 
$\overline{\rm MS}$ scheme, i.e. subtracting only the poles and the 
related constants in eq.(10), we would have been left with contributions 
which increase linearly with the gluino mass.} 
in such a way that it cancels exactly the mixing contributions $G_{mix}^i$
of eq.(11) [following Ref.\cite{R13}],
\beq
\delta\theta & = & \frac{1} {m_{\tilde{t}_1}^2-m_{\tilde{t}_2}^2} 
\left[ 
4 m_t \,m_{\tilde{g}} \,c_{2 \theta} \,B_0(m_{\tilde{t}_i}^2,m_t,
m_{\tilde{g}}) + c_{2 \theta} s_{2\theta} (A_0(m_{\tilde{t}_2}^2)- 
A_0(m_{\tilde{t}_1}^2) ) \right]
\eeq
In practice, this reduces to simply discarding the contributions 
of the diagrams Fig.2c. \s

The complete virtual corrections to the $t \ra \tilde{t}_i \chi$ 
decay width will be then given by
\begin{eqnarray}
\Gamma^V(t \rightarrow \tilde{t}_i\chi^0_j) & = &
\frac{\alpha}{3 \, m_t^3} \frac{\alpha_s}{4 \pi} 
       \;   \mbox{Re} \; \bigg\{
       ( m_t^2 - m_{\tilde{t}_i}^2 + m_{\chi}^2 ) 
                   (a_L^i \, G_L^i + a_R^i \, G_R^i) \nonumber \\ 
 & & \hspace{1.6cm} 
     + \;2 \, m_t \, m_{\chi} \epsilon_\chi \, 
                   ( a_L^i \, G_R^i + a_R^i \, G_L^i ) \bigg\} \,
    \lambda^{1/2}(m_t^2,m_{\chi}^2,m_{\tilde{t}_i}^2)
\end{eqnarray}
These corrections are infrared divergent; the divergencies are 
cancelled  after adding the real corrections. 

\subsubsection*{3.3 Real Corrections}

These corrections are obtained by adding the amplitudes of the diagrams
Fig.2e with real gluon emission and squaring the resulting amplitude;
the corrected decay width is then
\begin{eqnarray}
\Gamma_R^i & = & \frac{\alpha}{3m_t} \frac{\alpha_s}{\pi}
\bigg\{ 
  8 \; a_L^i \, a_R^i \; m_t \, m_{\chi} \epsilon_\chi \,  
 \big[\; ( m_{\chi}^2 - m_{\tilde{t}_i}^2
                                 - m_t^2 ) \, I_{01}
       -  m_{\tilde{t}_i}^2 \, I_{11}
       -  m_t^2 \, I_{00} 
       -  I_0  -  I_1  \big]
   \nonumber \\
 & & \hspace{1.6cm} +\; ({a_L^i}^2+{a_R^i}^2) \,
  \big[\;  2 \, ( m_{\tilde{t}_i}^2 - m_{\chi}^2  - m_t^2)
                      \, ( m_t^2 \, I_{00} + m_{\tilde{t}_i}^2 \, I_{11} 
                          + I_0 + I_1 )
   \nonumber \\ 
 & & \hspace{4.1cm}
       + 2 \, ( \; ( m_{\chi}^2 - m_{\tilde{t}_i}^2 )^2
                - m_t^4 ) \, I_{01}
       + I
       + I_0^1 \big]
\bigg\} 
\end{eqnarray}
where the phase space integrals $ I(m_t,m_{\tilde{t}_i},m_{\chi})
\equiv I $ are given by \cite{R14}
\beq
I_{00} & = & \frac{1}{4\,m_t^4}\bigg[ \kappa \, \log\bigg(
    \frac{\kappa^2}{\lambda\,m_t\,m_{\tilde{t}_i}\,m_{\chi}}\bigg)
    -\kappa-(m_{\tilde{t}_i}^2-m_{\chi}^2)\log
     \bigg(\frac{\beta_1}{\beta_2}\bigg)-m_t^2\,\log(\beta_0) \bigg] \non \\ 
I_{11} & = & \frac{1}{4\,m_{\tilde{t}_i}^2\,m_t^2}\bigg[ \kappa \, \log\bigg(
    \frac{\kappa^2}{\lambda\,m_t\,m_{\tilde{t}_i}\,m_{\chi}}\bigg)
    -\kappa-(m_t^2-m_{\chi}^2)\log
  \bigg(\frac{\beta_0}{\beta_2}\bigg)-m_{\tilde{t}_i}^2\,\log(\beta_1)
    \bigg] \non \\
I_{01} & = & \frac{1}{4\,m_t^2}\bigg[ -2\,\log\bigg(\frac{\lambda\,m_t\,
              m_{\tilde{t}_i}\,m_{\chi}}{\kappa^2} \bigg)\,\log(\beta_2) 
          + 2\,\log^2(\beta_2) - \log^2(\beta_0) - \log^2(\beta_1) \non \\
    &  & + 2\,\mbox{Li}\,(1-\beta_2^2) - \mbox{Li}\,(1-\beta_0^2)
         - \mbox{Li}\,(1-\beta_1^2) \bigg] \non \\
I & = & \frac{1}{4\,m_t^2} \bigg[ \frac{\kappa}{2}(m_t^2
                     +m_{\tilde{t}_i}^2+m_{\chi}^2)
       +2\,m_t^2\,m_{\tilde{t}_i}^2\,\log(\beta_2)
       +2\,m_t^2\,m_{\chi}^2\,\log(\beta_1)
       +2\,m_{\tilde{t}_i}^2\,m_{\chi}^2\,\log(\beta_0) \bigg] \non \\
I_0 & = & \frac{1}{4\,m_t^2} \bigg[ -2\,m_{\tilde{t}_i}^2\,\log(\beta_2)
         -2\,m_{\chi}^2\,\log(\beta_1)-\kappa \bigg] \non \\
I_1 & = & \frac{1}{4\,m_t^2}\bigg[ -2\,m_t^2\,\log(\beta_2)
         -2\,m_{\chi}^2\,\log(\beta_0)-\kappa \bigg] \non \\
I_0^1 & = & \frac{1}{4\,m_t^2}\bigg[ m_{\tilde{t}_i}^4 \,\log(\beta_2)
          -m_{\chi}^2 \,(2\,m_t^2-2\,m_{\tilde{t}_i}^2
          +m_{\chi}^2) \, \log(\beta_1)
          -\frac{\kappa}{4}\,(m_{\tilde{t}_i}^2-3\,m_t^2+5\,m_{\chi}^2) \bigg]
\eeq
with $\kappa = \lambda^{1/2}(m_t^2,m_{\tilde{t}_i},m_{\chi})$ and
\begin{equation} 
\beta_0 = \frac{m_t^2-m_{\tilde{t}_i}^2-m_{\chi}^2+\kappa}
    {2\,m_{\tilde{t}_i}\,m_{\chi}},\;\;
\beta_1 = \frac{m_t^2-m_{\tilde{t}_i}^2+m_{\chi}^2-\kappa}
    {2\,m_t\,m_{\chi}},\;\;
\beta_2 = \frac{m_t^2+m_{\tilde{t}_i}^2-m_{\chi}^2-\kappa}
    {2\,m_t\,m_{\tilde{t}_i}} 
\end{equation}

\subsection*{4. Results and Discussions}

The ${\cal O} (\alpha_s)$ partial widths of the decays $t \ra
\tilde{t}_2 \chi$ normalized to their Born expressions, are shown in
Figs.3 as a function of the lightest stop mass for the scenarios
discussed in section 2. For the numerical analysis, we choose $m_t=180$
GeV for the top quark mass and for the strong coupling constant we take
the value $\alpha_s= 0.118$. In Fig.3a, the corrections are shown for
$M_2=\mu=150$ (solid), $200$ (dashed) and 250 GeV (dotted); they are
negative for relatively small values of the gluino mass [solid 
line, $m_{\tilde{g}} \sim 530$ GeV]  and positive
for larger values of $m_{\tilde{g}}$ [dotted line, $m_{\tilde{g}} \sim
900$ GeV]; for intermediate values of the gluino mass [$m_{\tilde{g}}
\sim 700$ GeV, dashed line], they vary between -1\% and +1\%. 
In Fig.3b $M_2$ is again fixed at $400$ GeV and $\mu = -100$
(solid) and $150$ GeV (dashed). The QCD corrections are small in
general, but they can reach the level of $15$ \% for some stop masses.
\s 

The QCD corrections to the top decays into the lightest stop 
squark and one of the four neutralinos, $t \ra \tilde{t} \chi_j^0$,
in the scenario where $M_2 = -\mu=50$ GeV are shown in Fig.3c. Here
again, the corrections are rather small being of the order of $2 - 12$ \%
in all cases. As in Figs.1a--c, the sum was computed if more than one 
neutralino contributes. \s

For $M_2=400$ GeV and hence for large gluino mass, $m_{\tilde{g}} \sim
1.4$ TeV, the QCD corrections are negative and increase in absolute
value, reaching the level of 12\% for light stop squarks. This means
that the gluino does not decouple for large $m_{\tilde{g}}$. There
is a logarithmic dependence of the correction on the gluino mass 
which is visualized in Fig.4, where the correction is plotted against 
the gluino
mass for $\mu=-100$ GeV and $m_{\tilde{t}}=50$ GeV. This logarithmic
behaviour is due to the wave function renormalization and is a
consequence of the breakdown of supersymmetry as discussed in
Ref.\cite{R15}, where a similiar feature appears in the decay of a
squark into a massless quark plus a photino. 

\subsection*{5. Conclusion}

We have calculated the ${\cal O}(\alpha_s)$ QCD
corrections to the top decay into the lightest stop and a 
neutralino in the minimal supersymmetric extension of the Standard Model. 
These corrections can be either positive or negative and increase
logarithmically with the gluino mass. For gluino masses below 1 TeV,
they are at most of the order of ten percent and therefore, well under
control. \s

In the case where the top quark is lighter than its scalar partners, the
reverse process $\tilde{t} \ra t \chi_0$ might occur. This decay,
together with the decay $\tilde{t} \ra b \chi^+$, would be the 
dominant decay mode of stop squarks if they are lighter than $m_t+
m_{\tilde{g}}$. The QCD corrections to these processes are under 
investigation \cite{R16}. 

\bigskip

\nn {\bf Acknowledgements}: \s

\nn We thank F. Borzumati, R. H\"opker, T. Plehn and P.M. Zerwas for discussions.

\vspace*{2.cm}

\nn \subsection*{Figure Captions}

\begin{itemize}

\item[{\bf Fig.~1:~}]
Branching ratios for top decays into the lightest stop and neutralinos 
as a function of the stop mass in the Born approximation; $\tb$ is fixed 
to $\tb=1.6$ and the top quark mass to $m_t =180$ GeV. 
(a): $M_2=\mu = 150$ (solid line), 200 (dashed line) and 250 GeV (dotted 
line); 
(b): $M_2 = 400$ GeV and $\mu = -100$ GeV (solid line) and 
$M_2 = 400$ GeV and $\mu = 150$ GeV (dashed line); 
(c): $M_2 =-\mu = 50$ GeV the thin solid, dashed, dotted and dashed--dotted 
lines correspond respectively to the branching ratios of $t \rightarrow 
\tilde{t}_2 \chi^0_j$ with $j=1$--$4$ (the neutralinos are ordered with 
increasing mass), the thick solid line corresponds to the sum over all 
contributing neutralinos.

\item[{\bf Fig.~2~:~}] 
Feynman diagrams relevant for the ${\cal O}(\alpha_s)$ QCD corrections 
to the decay of the top quark into a stop squark and a neutralino. 

\item[{\bf Fig.~3:~}]
Relative size of the ${\cal O}(\alpha_s)$ QCD corrections to the decay rate
$t \ra \tilde{t} \chi$ as a function of the lightest stop squark mass. We 
use the same inputs as in Figs.1.

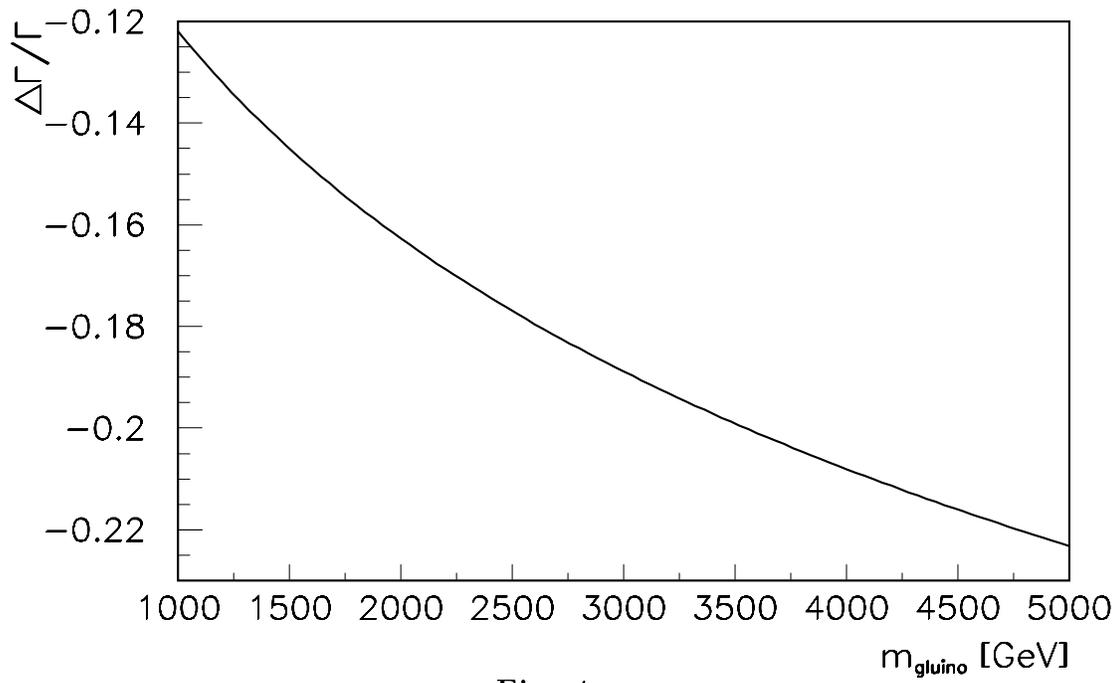
\item[{\bf Fig.~4:~}]
The size of the QCD corrections as a function of the gluino mass 
$m_{\tilde{g}}$ with $M_2 = 400$ GeV, $\mu=-100$ GeV and $m_{\tilde{t}_2}
\sim 50$ GeV.

\end{itemize}

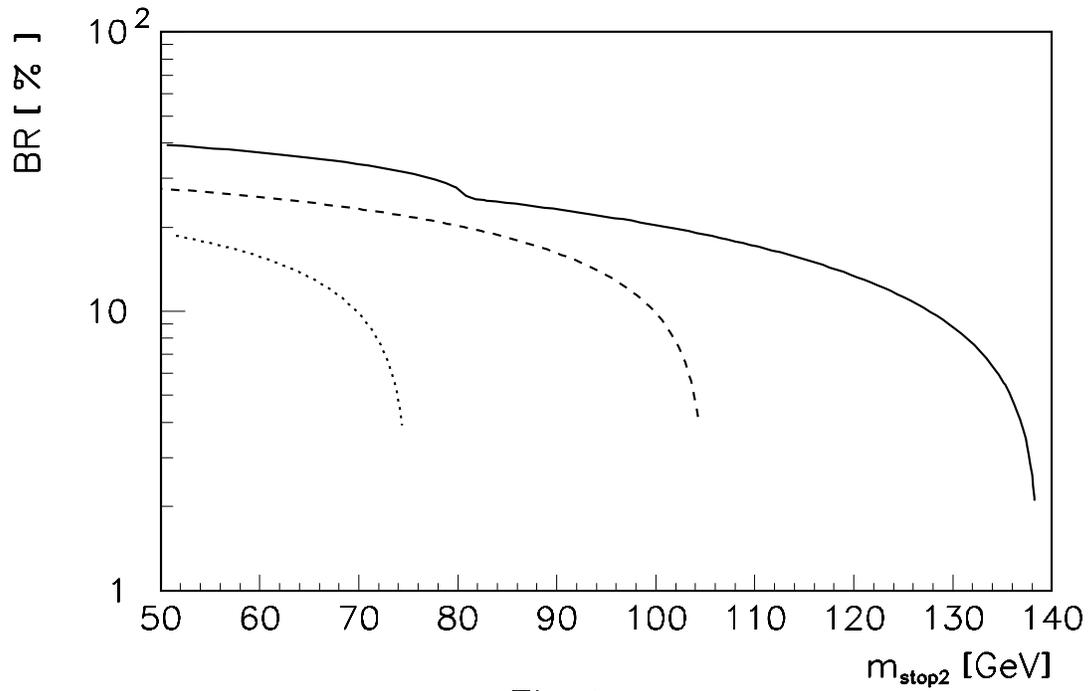
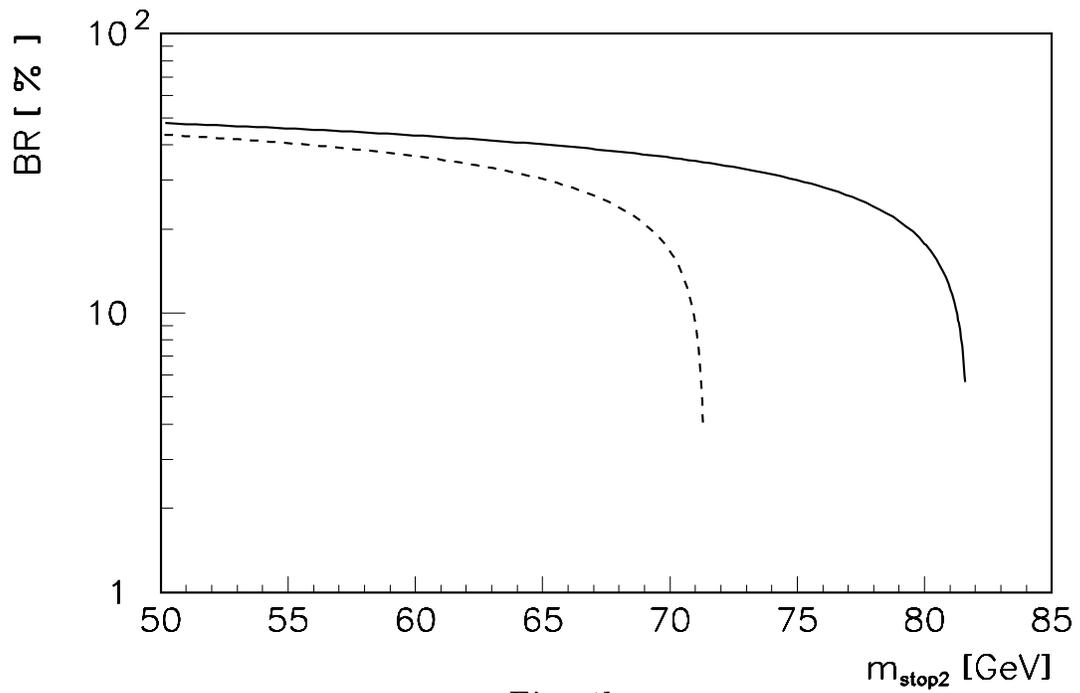
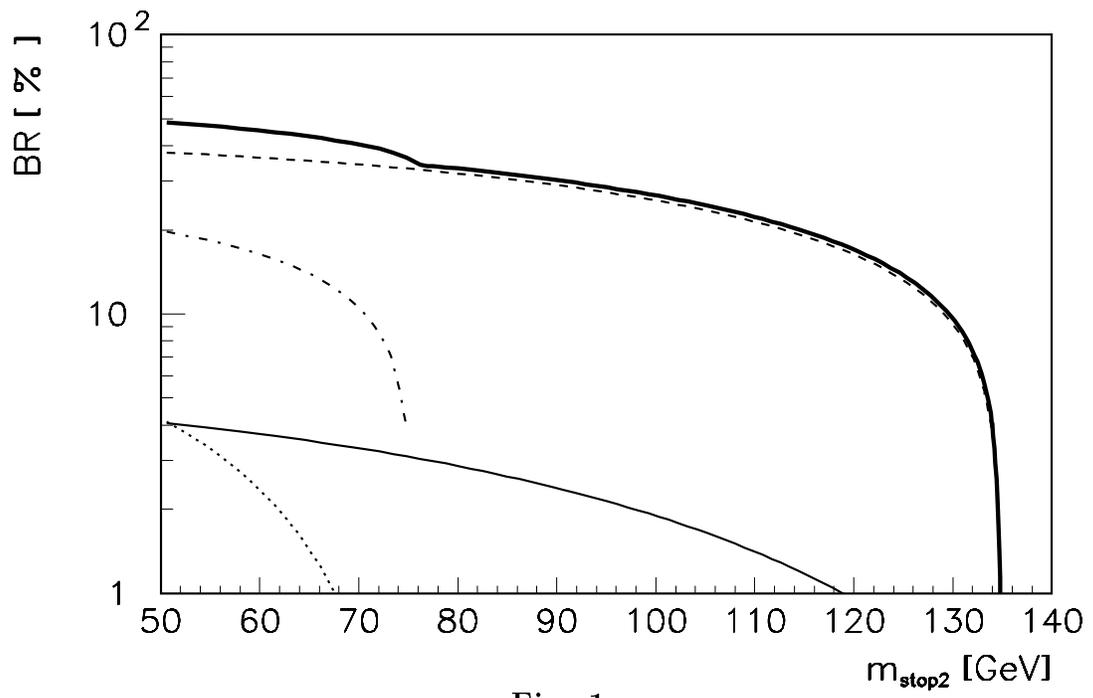
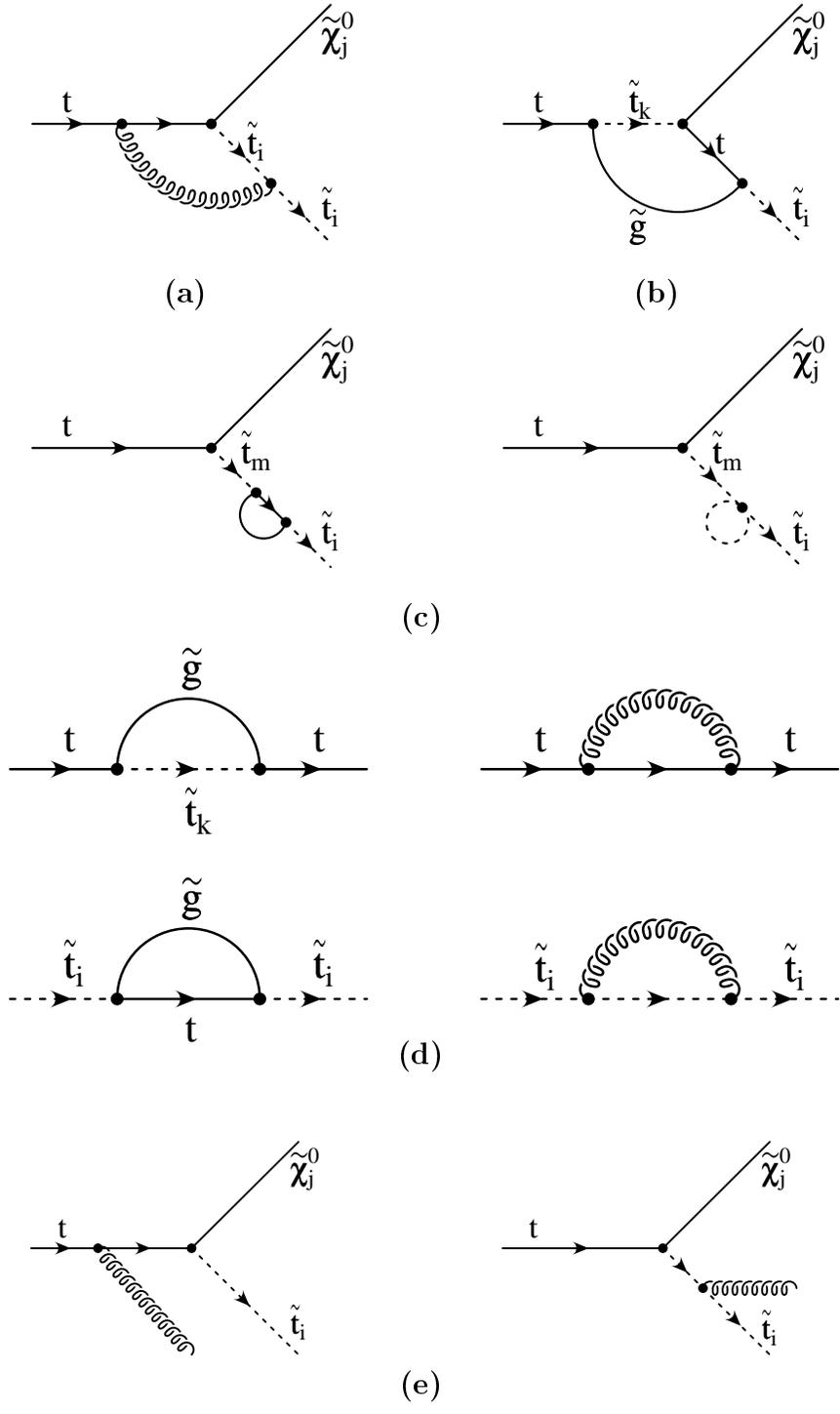
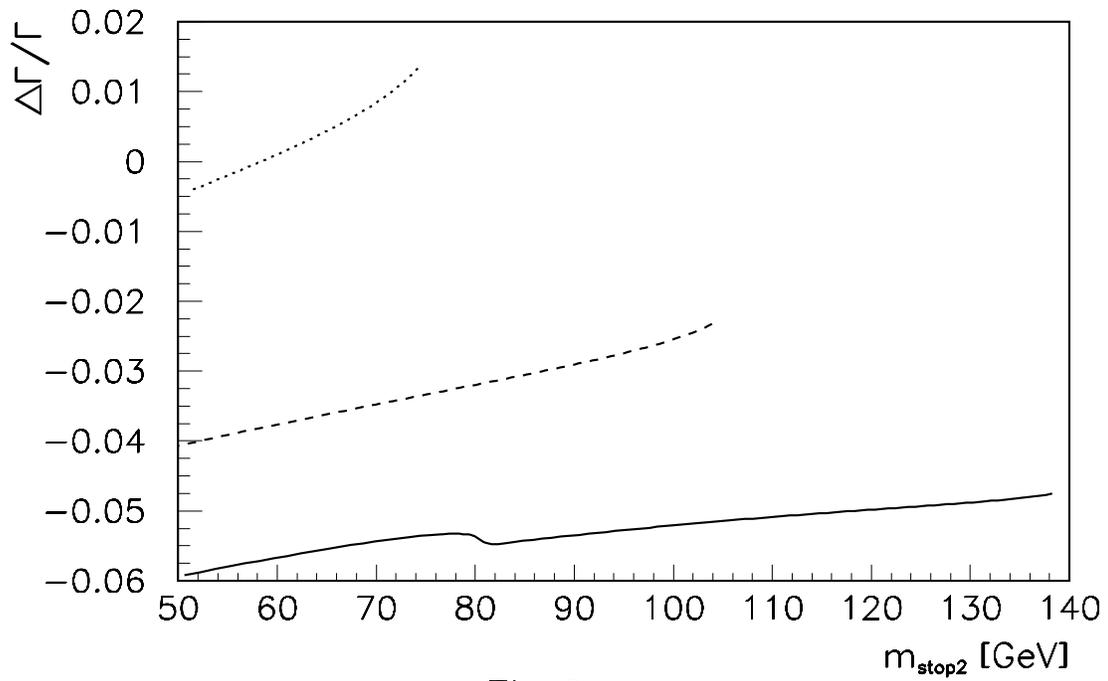
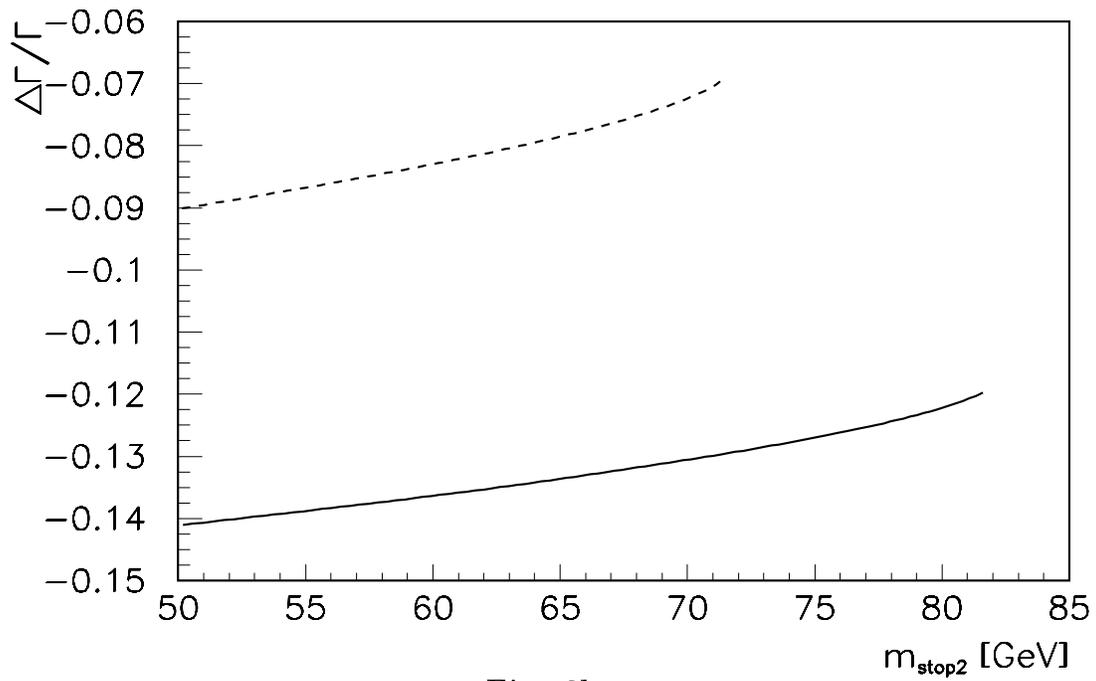
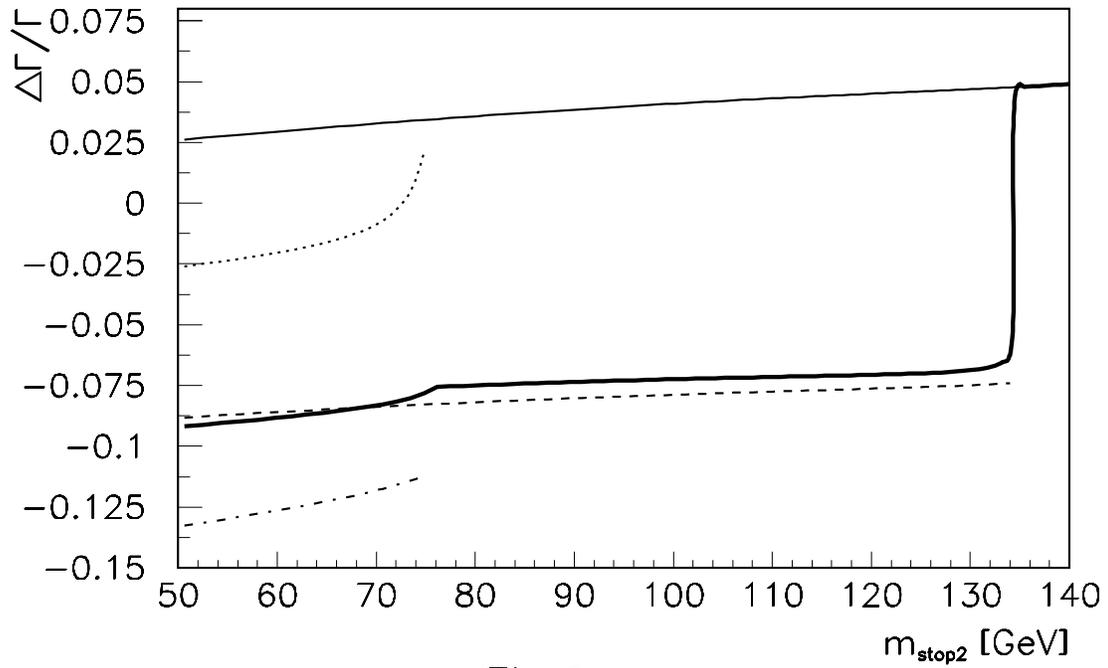
\end{document}